\documentstyle[12pt]{article}
\begin{document}
\baselineskip24pt
\title{The partition function of a 3-dimensional 
topological scalar-vector model}
\author{Bogus{\l}aw Broda\thanks{e-mail:
bobroda@krysia.uni.lodz.pl}
~and Ma{\l}gorzata Bakalarska\\
Department of Theoretical Physics, University of
{\L}\'od\'z\\
Pomorska 149/153, 90-236 {\L}\'od\'z, Poland}
\date{}
\maketitle

\begin{abstract}
A study of the partition function of a 3-dimensional 
scalar-vector model formally related via duality to the
Rozansky-Witten topological $\sigma$-model is presented.
The partition function is shown to consist of such 
topological quantities of a 3-dimensional manifold 
${\cal M}$ like a lattice sum, the Reidemeister-Ray-Singer 
torsion ${\tau}_{R} ({\cal M})$ and the Massey product.
\end{abstract}

\section{Introduction} 
A twisted (topological) version of 3d ${\cal N}=4$ SUSY 
$\sigma$-model with a hyper-K\"ahler manifold as a target 
space (RW model) has been analysed in detail by Rozansky and
Witten
in \cite{RozWit} (see, also \cite{Thom2} and \cite{MaMo}). 
In this letter, we will consider a 
scalar-vector (SV) $\sigma$-model with one variable dualized. More
precisely,
it consists of three scalar fields and one vector field 
which is dual to one out of four scalar fields appearing in
RW $\sigma$-model. Our SV model can be
interpreted as a variant related to low-energy version of 3d
${\cal N}=4$ SUSY $SU(2)$ gauge
model (Casson theory), or 3d ${\cal N}=4$ SUSY abelian one
with a matter
hypermultiplet (3d Seiberg-Witten theory), or as a
stand-alone model as well.

Let us consider a compact four-manifold  $X^4 = {\rm S}^1 
\times X$ with a product metric 
\[
\left(
\begin{array}{c|c}
g_{00} & 0\\ \hline
0 & g_{ij} ({\varphi}_1, {\varphi}_2, {\varphi}_3)
\end{array}
\right),
\]
as our target space. In 
this metric, we can perform a duality transformation 
for one scalar field, which replaces this field by a 
vector field \cite{HiKaLiRo} in RW action.
Since $X^4$ is not, in general, hyper-k\"ahlerian our SV 
model is not a priori topological.

We work on a 3-dimensional Euclidean manifold ${\cal M}$ and
denote local coordinates on ${\cal M}$ as
$x^{\mu}$, $\mu = 1, 2, 3$. ${\cal M}$ is endowed with a 
metric tensor $h_{\mu\nu}$. The bosonic scalar fields can be 
described as functions ${\varphi}^i$, $i=1, 2, 3$ 
with a metric tensor $g_{ij}$ on the target space $X$. 
The fermions are a scalar ${\eta}^I$ and 
a one-form ${\chi}^I_{\mu}$, where $I = 1, 2$.

Classical action of our model assumes the 
following form
\begin{eqnarray}
S & = & {\int}_{\cal M} \sqrt{h} d^3 x \left\{ \frac{1}{4} 
     F^{\mu\nu}F_{\mu\nu} + \frac{1}{2} g_{ij} 
     {\partial}_{\mu} {\varphi}^i {\partial}^{\mu} 
     {\varphi}^j + {\varepsilon}_{IJ} {\chi}^I_{\mu} 
     {\nabla}^{\mu} {\eta}^J + \mbox{} \right.\nonumber \\ 
     &&\left. \mbox{} + \frac{1}{2} \frac{1}{\sqrt{h}} 
     {\varepsilon}^{\mu\nu\rho} {\varepsilon}_{IJ} 
     {\chi}^I_{\mu} {\nabla}_{\nu} {\chi}^J_{\rho} +
     \frac{1}{6}
     \frac{1}{\sqrt{h}}{\varepsilon}^{\mu\nu\rho}
     {\Omega}_{IJKL} {\chi}^I_{\mu} {\chi}^J_{\nu} 
     {\chi}^K_{\rho} {\eta}^L \right\},       
\label{1}
\end{eqnarray}
where $F_{\mu\nu} = {\partial}_{\mu} A_{\nu} - 
{\partial}_{\nu} A_{\mu}$ is the usual $U(1)$ gauge 
field strength. The ``curvature'' tensor ${\Omega}_{IJKL}$ 
is a completely symmetric tensor field on $X$.
The covariant derivative of fermions, here denoted as 
${\nabla}_{\mu}$, is defined using the pullback of the Levi-Civita 
connection on $X$,
\begin{equation}
{\nabla}_{\mu} = {\partial}_{\mu} {\delta}^I_J + 
({\partial}_{\mu} {\varphi}^i) {\Gamma}^I_{iJ}.
\label{2}
\end{equation}

\section{The classical contribution}       

Let us consider classical contribution to the partition 
function coming from $U(1)$ gauge field $A_{\mu}$, 
i.e. the contribution of classical saddle-points 
\cite{Ver}, \cite{Wit5}
\begin{equation}
Z_{cl} = \sum_{\begin{array}{c}\mbox{\scriptsize saddle}\\[-
3mm]
\mbox{\scriptsize points} \end{array}} e^{- S [A_{cl}]}.
\end{equation}
When the three-manifold ${\cal M}$ has non-trivial homology 
two-cycles ${\Sigma}_I$, i.e. closed surfaces that are not 
boundaries, there exist field configurations with 
non-zero flux through these surfaces, that must obey 
generalized Dirac quantization condition
\begin{equation}
    {\int}_{\Sigma_I} F = 2 \pi m^I, \qquad m^I \in Z,
\end{equation}
where $I=1, \ldots, b_2({\cal M})(={\rm dim} H_2 ({\cal
M})$,
the second Betti number).
This tells us that in the absence of sources $F$ 
can be written as
\begin{equation}
    F= 2 \pi \sum_I m^I {\alpha}_I,  
\end{equation}
where ${\alpha}_I$ is an integral basis of harmonic 
2-forms, which by definition satisfy $d {\alpha}_I= 
d*{\alpha}_I=0$ and are normalized so that 
${\int}_{\Sigma_I} {\alpha}_J = {\delta}^I_J$.  
The classical saddle-points are labelled by the integer 
magnetic fluxes $m^{I}$. The classical action for this 
field configuration is   
\begin{equation}
    S[m^I]= {\pi}^2 \sum_{I,J} m^I G_{IJ} m^J,
\end{equation}
where
\begin{equation}
G_{IJ} = {\int}_{\cal M} {\alpha}_I \wedge *{\alpha}_J, 
\end{equation}
represents the metric on the space of harmonic two-forms. 
Finally, 
\begin{equation}
   Z_{cl} = \sum_{\scriptsize m^I} e^{-S[m^I]}.
\end{equation}

The zero modes of the vector field are tangent to the space 
of classical minima, which is a torus of dimension 
$b_1({\cal M})$. The classical part should involve also 
integration over the $b_1$-torus.

The minima of the action corresponding to the scalar 
field ${\varphi}^i$ are the constant maps of ${\cal M}$ 
to $X$. So we will expand around those according to 
\cite{RozWit}. To take into account the bosonic zero modes 
one must introduce 
``collective coordinates'' and integrate over the space 
of all constant maps of ${\cal M}$ to $X$. Thus, we split 
the bosonic field ${\varphi}^i$ into a sum of a constant 
and fluctuating part,
\begin{equation}
     {\varphi}^i_0 + {\varphi}^{i} (x). 
\end{equation}
We define a partition function $Z_{X} ({\cal M}; 
{\varphi}^i_0)$ of fixed ${\varphi}^i_0$, and obtain 
the partition function $Z_{X} ({\cal M})$ 
as an integral over the three-dimensional target space $X$
\begin{equation}
         Z_{X} ({\cal M}) = \frac{1}{(2\pi\hbar)^{3/2}}
                          {\int}_{X} Z({\cal M}; {\varphi}^
                          i_0)
                         \sqrt{g} d^3 {\varphi}^i_0,
\end{equation}
where $Z({\cal M}; {\varphi}^i_0)$ is a product of two
factors
\begin{equation}
       Z({\cal M}; {\varphi}^i_0) = Z_0 ({\cal M}; {\varphi}
       ^i_0)
                  Z_{\eta\chi\varphi} ({\cal M}, X;
                  {\varphi}^i_0).
\end{equation}
Here $Z_0 ({\cal M}; {\varphi}^i_0)$ is the 1-loop 
contribution of non-zero modes of $\varphi$ and $A$, while
$Z_{\eta\chi\varphi}
({\cal M}, X; {\varphi}^i_0)$ is the exponential of the sum 
of all Feynman diagrams of two or more loops, in the 
background field of given ${\varphi}^i_0$.

\section{The one-loop contribution}    

Let us first determine the one-loop contribution 
$Z_{0}({\cal M}; {\varphi}^i_0)$. We work with the part 
of the action which is quadratic in the vector field 
$A_{\mu}(x)$, in fluctuating bosonic fields 
${\varphi}^i (x)$ and in fermionic fields 
${\eta}^I(x)$, ${\chi}^I_{\mu}(x)$ 
\begin{eqnarray}
S_{0} &=& {\int}_{\cal M} d^3 x \sqrt{h} \left\{ \frac{1}{4}
     F^{\mu\nu} F_{\mu\nu} + \frac{1}{2} g_{ij} {\partial}_{
     \mu}
     {\varphi}^i {\partial}^{\mu} {\varphi}^j +
     {\varepsilon}_{IJ}
     {\chi}_{\mu}^I {\nabla}^{\mu} {\eta}^J + \mbox{}\right.
     \nonumber \\
     &&\left. \mbox{} + \frac{1}{2} \frac{1}{\sqrt{h}}
     {\varepsilon}^{\mu\nu\rho} {\varepsilon}_{IJ} {\chi}_{
     \mu}^I
     {\nabla}_{\nu} {\chi}_{\rho}^J \right\}.
  \label{3.1}
\end{eqnarray}
(The tensors $g_{ij}$, ${\varepsilon}_{IJ}$ and implicit 
Christoffel symbols ${\Gamma}^i_{jk}$ are 
taken at the point ${\varphi}^i_0$ of $X$).
Gauge invariance of the action requires gauge fixing 
and introduction of the Faddeev-Popov ghost fields 
$c$, $\bar{c}$
\begin{equation}
  S_{gauge}= {\int}_{\cal M} d^3 x \sqrt{h} \left\{\frac{1}{
  2} ({\nabla}^{\mu} {A}_{\mu})^2 +
                    {\partial}^{\mu} \bar{c}
                    {\partial}_{\mu} c \right\}.
  \label{3.2}     
\end{equation}
Supplementing the action (\ref{3.1}) with (\ref{3.2}), 
we obtain
\begin{eqnarray}
S^{\prime}_{0} &=& {\int}_{\cal M} d^3 x \sqrt{h} \left\{ 
             \frac{1}{4} F^{\mu\nu} F_{\mu\nu} + \frac{1}{2} 
             g_{ij} {\partial}_{\mu} {\varphi}^i {\partial}_
             {\mu}
             {\varphi}^j + {\varepsilon}_{IJ} {\chi}^I_{\mu} 
             {\nabla}^{\mu}{\eta}^J + \mbox{}\right.
             \nonumber \\
             &&\left. \mbox{} +  \frac{1}{2} 
             \frac{1}{\sqrt{h}} {\varepsilon}^{\mu\nu\rho} 
             {\varepsilon}_{IJ} {\chi}^I_{\mu}
             {\nabla}_{\nu}
             {\chi}^J_{\rho} + \frac{1}{2} ({\nabla}^{\mu} 
             {A}_{\mu})^2 + {\partial}^{\mu} \bar{c} 
             {\partial}_{\mu} c \right\}.
\label{3.3}
\end{eqnarray}

It will appear that the partition function $Z_0$
corresponding
to (\ref{3.3}) essentially consists of the 
Reidemeister-Ray-Singer torsion of 3d manifold ${\cal M}$. 
Namely,
\begin{equation}
        Z_{0} ({\cal M}; {\varphi}^i_0) = \int [{\cal D}Y] 
        {\rm exp} \left\{ - S^{\prime}_0 [Y]\right\},
 \end{equation}
where the integration measure $[{\cal D}Y]$ is taken over
all
the fields: 
$A_{\mu}$, ${\varphi}^i$, ${\eta}^I$, ${\chi}_{\mu}^I$, $c$,
$\bar{c}$.
The path integral of the scalar bosonic and ghost fields
gives a
net contribution
\begin{equation}
      ({\rm det}^{\prime} (-{\Delta}_0))^{-\frac{1}{2}},
\end{equation}
where ${\Delta}_i = {\nabla}^{\mu} {\nabla}_{\mu}$ ($i=0,1$)
is
a laplacian acting on $i$-forms on ${\cal M}$ and the prime
 means that we exclude zero modes.
Now let us introduce an operator $L_{-}$ which acts on the 
direct sum of zero- and one-forms on ${\cal M}$
\cite{RozWit}
\begin{equation}
     L_{-}(\eta, {\chi}_{\mu}) = \left(-{\nabla}^{\mu}
     {\chi}_{\mu},
     {\nabla}_{\mu} \eta  + h_{\mu\nu} \frac{1}{\sqrt{h}} 
     {\varepsilon}^{\nu\rho\lambda} {\partial}_{\rho} 
     {\chi}_{\lambda} \right).
\end{equation} 
Then the fermionic part of the action (\ref{3.3}) becomes 
a quadratic form
\begin{equation}
      \frac{1}{2} {\varepsilon}_{IJ} \left< {\eta}^I, 
      {\chi}_{\mu}^I \mid L_{-} \mid {\eta}^J, 
      {\chi}_{\mu}^J \right>.  
\end{equation}
The fermionic one-loop contribution with zero 
modes removed is \cite{Wit4}
\begin{equation}
         {\rm det}^{\prime} L_{-}.
\end{equation}
Finally, the integration over the gauge field $A_{\mu}$
yields
\begin{equation}
         ({\rm det}^{\prime} ( -
         {\Delta}_{1}))^{-\frac{1}{2}},
\end{equation}
so that the total one-loop contribution of non-zero modes is 
\begin{equation}
         Z_0 ({\cal M}; {\varphi}^i_0)= \frac{{\rm det}^{
         \prime}
         L_{-}}{({\rm det}^{\prime}
         ( - {\Delta}_{0}))^{\frac{1}{2}} ({\rm
         det}^{\prime}
         ( - {\Delta}_{1}))^{\frac{1}{2}}}.
  \label{3.4}
\end{equation}
The absolute value of the ratio of the determinants in 
(\ref{3.4}) is related to the Reidemeister-Ray-Singer 
analytic torsion ${\tau}_R ({\cal M})$ \cite{RozWit}, 
\cite{Roz1} and \cite{Roz2}
\begin{equation}
\left | \frac{{\rm det}^{\prime} L_{-}}{({\rm det}^{\prime}
    (- {\Delta}_{0}))^{\frac{1}{2}} ({\rm det}^{\prime}
    (- {\Delta}_{1}))^{\frac{1}{2}}} 
    \right | = {\tau}_{R}^{-2} ({\cal M}). 
\end{equation} 

\section{Zero modes and propagators}      

The partition function is plagued by zero modes, which we 
have temporarily removed by hand. There are the following
four sorts
of zero modes:
\begin{list}{}{}
\item[(1)] 3 scalar boson zero modes corresponding to 
${\varphi}^i$; 
\item[(2)] $b_1$ vector boson zero modes of $A_{\mu}$;
\item[(3)] 1 ghost zero mode for $c$, $\bar{c}$;
\item[(4)] 2 scalar fermion zero modes of ${\eta}^I$ and 
$2 b_1$ one-form fermion zero modes of ${\chi}_{\mu}^I$.  
\end{list}
The ghost zero mode can be removed instantaneously without 
any consequences because it should not be present in the 
partition function from the very beginning at all, as 
the gauge transformation corresponding to the constant 
(zero) mode acts trivially on $A_{\mu}$.  The rest of 
boson and fermion zero modes have been shifted from the 
one-loop calculation as they would produce trivial
infinities
and zeros in the partition function respectively. 
Actually, the boson zero modes have been already dealt
with---${\varphi}^i_0$-integration for ${\varphi}$ and
$b_1 ({\cal M})$-torus integration for $A$. The fermion zero
modes
will saturate higher-order loops. 

The propagators in our model are of the following form:
\begin{equation}
  \left< {\varphi}^i (x) {\varphi}^j (y) \right> = 
   - \hbar g^{ij} G^{\prime (0)}(x, y),
\end{equation}

\begin{equation}
  \left< {\chi}^I_{\mu} (x) {\eta}^J (y) \right> = \hbar
  {\varepsilon}^{IJ} {\partial}_{\mu} G^{\prime (0)}(x, y),
\end{equation}

\begin{equation}
  \left< {\chi}^I_{\mu} (x) {\chi}^J_{\nu} (y) \right> 
\equiv \hbar G^{(\chi)}_{\mu\nu} (x, y) =
  \frac{1}{2} \hbar {\varepsilon}^{IJ} h^{- 1/2} 
  h_{\mu\lambda} {\varepsilon}^{\lambda\kappa\rho} 
    {\partial}_{\rho} G^{\prime (1)}_{\kappa\nu} (x, y),
\label{4.1}
\end{equation}
where $G^{\prime (i)} (x, y)$ is an inverted Laplacian for 
$i$-forms  with zero modes removed.
 
\section{Feynman diagrams}     

Let us analyse higher-order perturbative calculation, 
thus the diagrams that have a chance to absorb the fermionic
zero modes.
We may limit our attention to only those Feynman diagrams 
(analogously to \cite{RozWit}), whose contribution is of 
order ${\hbar}^s$, because the rest of diagrams is 
equal to zero. 

Let us consider a diagram with $V$ vertices,
$V=V_0+V_1+V_2+V_3$,
where $V_n$ ($n=0,1,2,3$) means the number of vertices with
$n$ (fermionic) legs of type ${\chi}_{\mu}$.
Let $L$ be the total number of legs. Each vertex in the 
diagram carries a factor of ${\hbar}^{-1}$, so the 
vertices taken together bring a factor ${\hbar}^{-V}$. 
Each propagator carries a factor of $\hbar$, and each 
fermionic zero mode carries a normalization factor 
of ${\hbar}^{\frac{1}{2}}$. Therefore the total 
contribution of propagators and external legs is 
${\hbar}^{\frac{L}{2}}$. So we get
\begin{equation}
\frac{L}{2} - V = s,
\end{equation}
where
\begin{equation}
      s = \frac{3}{2} + \frac{b_1}{2} - \frac{1}{2} = 
      \frac{b_1}{2} + 1,
\end{equation}
which follows from the previous section.
To narrow searching procedure it is also necessary to take
into account
some inequalities among numbers of vertices:
\begin{eqnarray}
&& L \geq 3V_0 + 4(V_1 + V_2 + V_3), \nonumber\\
&& V_1 + V_3 \geq 2, \\
&& V_1 + 2V_2 + 3V_3 \geq 2b_1,\nonumber
\end{eqnarray}
which, in turn, imply the following 
conditions for the Feynman diagrams:
\begin{eqnarray*}
&& 1 \geq V_0 + V_1 + V_2,\\
&& V_3 \geq 2V_0 + V_1 + 2V_2,\\
&& 4 \geq 2V_0 + 3V_1 + 2V_2 + V_3.
\end{eqnarray*}

This set of inequalities yields exactly seven 
different solutions (types of Feynman diagrams). Six 
 of them vanish because of parity symmetry, or 
because of geometrical identities for (harmonic) one-forms:
$\omega \wedge \omega = 0$, and $d \omega = 0$.
The only exception is the Feynman diagram 
(surviving for $b_1 ({\cal M}) = 2$) with $V_3=2$, where the
vertices are connected by a single
$\left< \chi \chi \right>$ propagator (all 
remaining legs absorb the zero modes of ${\eta}^I$ 
and ${\chi}^I_{\mu}$)

{\renewcommand{\arraystretch}{0.6}
\setlength{\unitlength}{1mm}
\begin{picture}(140,45)(0,0)
\put(30,15){\circle*{1}}
\multiput(30,15)(0,2){7}{\line(0,1){1}}
\multiput(30,15)(0,-2){7}{\line(0,-1){1}}
\multiput(30,15)(-2,0){8}{$.$}
\multiput(30,15)(2,0){18}{\line(1,0){1}}
\put(31,10.5){$\rm V_3$}

\put(100,15){\circle*{1}}
\multiput(100,15)(-2,0){18}{\line(-1,0){1}}
\multiput(100,15)(0,2){7}{\line(0,1){1}}
\multiput(100,15)(0,-2){7}{\line(0,-1){1}}
\multiput(100,15)(2,0){8}{$.$}
\put(101,10.5){$\rm V_3$}
\put(60,18){$\left< \chi\chi \right>$}
\end{picture}

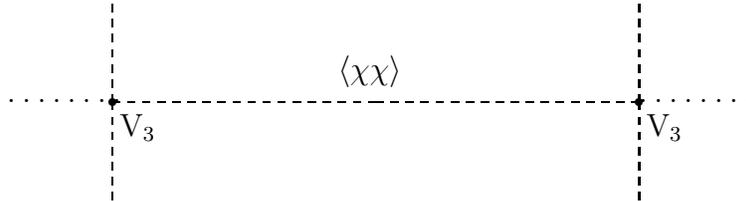
{\bf Fig. 1} The only non-vanishing higher-order Feynman 
diagram, giving rise to the Massey product.


The  integral corresponding to this diagram
\begin{equation}
     I({\cal M}) = {\int}_{\cal M} {\varepsilon}^
     {{\mu}_1 {\mu}_2 {\mu}_3} {\varepsilon}^{{\nu}_1 
     {\nu}_2 {\nu}_3} {\omega}^{(1)}_{{\mu}_1} (x) 
     {\omega}^{(1)}_{{\nu}_1} (y) {\omega}^{(2)}_{{\mu}_2} 
     (x) {\omega}^{(2)}_{{\nu}_2}(y) G^{(\chi)}_{{\mu}_3 
     {\nu}_3} (x, y) d^3 x d^3 y,
\end{equation}
can be evaluated following  \cite{RozWit}, showing 
the appearance of the Massey product.
The weight function is equal to 
\begin{equation}
 a(X) = \frac{1}{{(2\pi)}^{3/2}} {\int}_X \sqrt{g} d^3 
 {\varphi}^i_0 {\varepsilon}^{I_1 J_1} {\varepsilon}^{I_2
 J_2}
 {\varepsilon}^{I_3 J_3} {\varepsilon}^{I_4 J_4} 
 {\Omega}_{I_1 I_2 I_3 I_4} {\Omega}_{J_1 J_2 J_3 J_4}.  
\end{equation}
So the contribution of the above Feynman diagram to 
the partition function of our scalar-vector model is equal 
\begin{equation}
        Z_{\eta\chi\varphi} ({\cal M}, X) =  \frac{1}{2} 
        a(X) I({\cal M}).
\end{equation}
Finally, the partition function of the scalar-vector model 
has the following form
\begin{equation}
Z_X ({\cal M}) = \frac{1}{2} Z_0 ({\cal M}) a(X) I({\cal
M}),
\end{equation}
where $Z_0 ({\cal M})$ is given by eq.(\ref{3.4}).

\section{Summary}

For SV  $\sigma$-model 
the partition function has been exactly calculated on the  
three-dimensional manifold ${\cal M}$ giving the 
following topological quantities: a lattice sum 
containing the metric on the space of harmonic two-forms, 
the Reidemeister-Ray-Singer torsion ${\tau}_R ({\cal M})$ 
and the Massey product. The lattice sum is of classical 
origin, whereas the torsion is coming from the one-loop 
contribution (functional determinants without zero modes). 
Almost all ``higher-order loops'' are killed by fermion zero 
modes, and the only one Feynman diagram surviving  (for
$b_1 ({\cal M}) = 2$) yields the Massey product.

We should stress that SV $\sigma$-model discussed in the
present paper is not equivalent to RW model. First of all,
although SV action can be obtained from
RW action via duality, the duality performed, as a purely
local operation transfers no global/topological information
between the models. Besides, the hyper-k\"ahlerian condition
is relaxed in our case, and a priori the model is not topological.
Furthermore, a posteriori, the non-equivalence of both
 perturbative calculi is visible. 

\section*{Acknowledgements}

The paper has been supported by KBN grant 2P03B08415 and
partially by U\L~grant 248.
The authors would like to thank  Guowu Meng  for a critical remark.


\begin{thebibliography}{99}

\bibitem{RozWit}
L.~Rozansky and E.~Witten, ``Hyper-K\"ahler Geometry and 
Invariants of Three-Manifolds'', E-print hep-th/9612216.

\bibitem{Thom2}
G.~Thompson, ``On the Generalized Casson Invariant'', 
E-print hep-th/9811199.

\bibitem{MaMo}
M.~Marino and G.~Moore, ``Three-manifold topology and the 
Donaldson-Witten partition function'', Nucl. Phys. B 547,
569
(1999).

\bibitem{HiKaLiRo}
N.~J.~Hitchin, A.~Karlhede, U.~Lindstr\"om and M.~Ro\v{c}ek,
Commun.\ Math.\ Phys. 108, 535 (1987).

\bibitem{Ver}
E.~Verlinde, Nucl. Phys. B 455, 211 (1995).

\bibitem{Wit5}
E.~Witten, ``On S-Duality In Abelian Gauge Theory'', 
E-print hep-th/9505186.

\bibitem{Wit4}
E.~Witten, Commun.\ Math.\ Phys. 121, 351 (1989).

\bibitem{Roz1}
L.~Rozansky, Commun.\ Math.\ Phys. 171, 279 (1995).

\bibitem{Roz2}
L.~Rozansky, ``Witten's invariant of 3-dimensional 
manifolds: loop expan\-sion and surgery calculus'', 
Knots and Applications, ed.~L.~H.~Kauffman, 271--299.

\end{thebibliography}
\end{document}